\documentclass[aps,prb,twocolumn,showpacs]{revtex4}
\usepackage{epsfig}
\usepackage{hyperref}

\begin{document}
\draft 
\def\il{I_{low}} 
\def\iu{I_{up}} 
\def\eeq{\end{equation}}
\def\ie{i.e.}  
\def\etal{{\it et al. }}  
\def\prb{Phys. Rev. {\bf B}}
\def\pra{Phys. Rev. {\bf A}} 
\def\prl{Phys. Rev. Lett. }
\def\pla{Phys. Lett. A } 
\def\pb{Physica B}
\def\ajp{Am. J. Phys. }  
\def\mpl{Mod. Phys. Lett. {\bf B}} 
\def\ijmp{Int. J. Mod. Phys. {\bf B}} 
\def\ijp{Ind. J. Phys. }
\def\ijpap{Ind. J. Pure Appl. Phys. }
\def\ibmjrd{IBM J. Res. Dev. }
\def\pjp{Pramana J. Phys.}

\title{Current magnification effect in mesoscopic systems at
  equilibrium}

\author{Colin Benjamin} 
\email{colin@iopb.res.in}
\author{A. M. Jayannavar}
\email{jayan@iopb.res.in}
\address{Institute of Physics, Sachivalaya Marg, Bhubaneswar 751 005,
  Orissa, India}

\date{\today}

\begin{abstract}
  We study the current magnification effect and associated circulating
  currents in mesoscopic systems at equilibrium. Earlier studies have
  revealed that in the presence of transport current (non-equilibrium
  situation), circulating currents can flow in a ring even in the
  absence of magnetic field. This was attributed to current
  magnification which is quantum mechanical in origin.  We have shown
  that the same effect can be obtained in equilibrium systems,
  however, in the presence of magnetic flux. For this we have
  considered an one-dimensional open mesoscopic ring connected to a
  bubble, and the system is in contact with a single reservoir.  We
  have considered a special case where bubble does not enclose
  magnetic flux, yet circulating currents can flow in it due to
  current magnification.
\end{abstract}


\maketitle

Mesoscopic physics deals with the realm which is in between the
microscopic (atomic or molecular) scale and macroscopic one. In these
systems quantum phase coherence length $L_{\phi}$ exceeds the sample
size $L$. These systems have provided several, often counter-intuitive
new results exploring truly quantum effects beyond the atomic
realm\cite{dephase:imry,dephase:psd}.  These systems are expected to
reveal the crossover between quantum and the macroscopic classical
regimes, which is of fundamental interest. The notion of intrinsic
decoherence and dephasing of a particle interacting with its
environment are being actively pursued and experimentally being
analysed\cite{dephase:imry,dephase:sprinzak}. The decoherence
mechanism signals the limit beyond which the system dynamics
approaches the classical behavior. One of the prominent mesoscopic
effect is that of observation of persistent currents in metallic rings
enclosing magnetic flux.  B\"{u}ttiker, Imry and Landauer
predicted\cite{bil} the existence of equilibrium persistent current in
an ideal one-dimensional metallic ring in presence of magnetic flux,
with a period of $\phi_{0}$, $\phi_{0}$ being the elementary flux
quanta $hc/e$. The existence of persistent currents have been verified
experimentally\cite{levy}. Persistent currents occur in both open and
isolated closed systems
\cite{cheung2,buti,coupled,deo_mplx,prb1994,physicapareek}.  Since
then circulating currents have been predicted in open systems in
presence of a transport current. This phenomenon is associated with
current magnification effect in mesoscopic
rings\cite{prb1994,physicapareek,colin_cm}. For this we consider a
metallic loop connected to two reservoirs by two ideal leads.
Transport current $I$ flows through the system when the two reservoirs
are kept at different chemical potentials, say $\mu_{1}$ and $\mu_{2}$
respectively. The upper and lower arms of the ring are of different
lengths and currents $I_{1}$ and $I_{2}$ flow in these such that
$I_{1}$ $\neq$ $I_{2}$. The basic law of current conservation namely,
Kirchoff's law demands that $I=I_{1}+I_{2}$. In the classical case
both $I_{1} $ and $I_{2}$ are positive and flow along the direction of
the applied chemical potential. However, when quantum mechanically
currents are calculated depending upon the length parameters it is
found that for particular values of Fermi energy $I_{1}$ (or $I_{2}$)
can be much larger than $I$. Current conservation thus dictates
$I_{2}$ (or $I_{1}$) to be negative such that $I=I_{1}+I_{2}$. The
property that current in one of the arms is larger than the transport
current is referred to as \emph{current magnification} effect. This
quantum effect has no classical analog in equilibrium. In such a
situation one can interpret that the negative current flowing in one
arm continues to flow as a circulating current in the
loop\cite{prb1994,physicapareek,colin_cm}.  Our procedure of assigning
a circulating current is exactly the same as the procedure well known
in classical LCR ac network analysis.  When a parallel resonant
circuit(capacitance C connected in parallel with a combination of
inductance L and resistance R) is driven by external electromotive
force(generator), circulating currents arise in the circuit at
resonant frequency\cite{book}.  The magnitude of the negative current
in one of the arms flowing against the direction of the applied
current is taken to be that of the circulating current.  When the
negative current flows in the upper arm the circulating current
direction is taken to be anticlockwise (or negative) and when it flows
in the lower arm the circulating current direction is taken to be
clockwise(or positive)\cite{prb1994,physicapareek,colin_cm,book}.
  
It should be noted that these circulating currents arise in the
absence of magnetic flux and in presence of transport currents (i.e.,
in a non-equilibrium system).  It has also been shown that impurities
affect current magnification in a non-trivial way. In fact, impurities
can enhance current magnification as opposed to the conventional
wisdom that impurities would degrade current
magnification\cite{prb1994,colin_cm}.  Studies on circulating currents
in mesoscopic open rings have been extended to thermal
currents\cite{mosk} and to spin currents in the presence of
Aharonov-Casher flux\cite{choi}.  Recently this effect has been
studied in presence of spin-flip scattering which causes dephasing of
electronic motion\cite{colin_cm,joshi}.
  
In the present work we are interested in the basic question, whether
current magnification can occur in equilibrium systems. For this we
consider the system as depicted in Figure (1). The static localised flux
piercing the loop is necessary to break the time reversal symmetry and
induce a persistent current in the system. The geometry we consider is
a one-dimensional ring coupled to a bubble. The system is connected to
a reservoir at chemical potential $\mu$. The reservoir acts as an
inelastic scatterer and as a source of energy dissipation\cite{buti}.
We would like to emphasize that the magnetic flux is localised in a
finite region. The loops J1J2aJ3J1 and J1J2bJ3J1 enclose the localised
flux $\phi$. However, the bubble J2aJ3bJ2 does not enclose the flux
$\phi$. The special situation we have considered, is to answer the
question of existence of circulating currents in equilibrium systems.
We show that circulating currents(due to current magnification) arise
in a bubble which does not enclose a magnetic flux. We would like to
mention here that the current magnification effect and the associated
circulating currents arise even when the magnetic field extends over
the entire sample. However, for this the treatment is involved as one
has to study separately persistent as well as circulating currents in
the bubble as they have different symmetry properties. This has been
studied in a simple loop in the presence of both transport currents
and magnetic flux\cite{physicapareek}.

\begin{figure}
\protect\centerline{\epsfxsize=3.5in\epsfbox{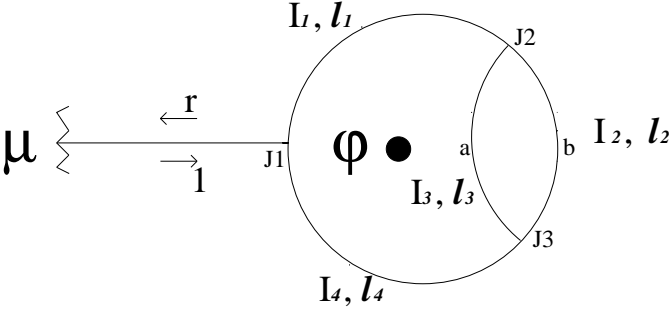}}
\caption{One dimensional mesoscopic ring coupled to a bubble with a
  lead connected to a reservoir at chemical potential $\mu$. The
  localised flux $\phi$ penetrates the ring.}
\end{figure}
  
In the local coordinate system the wavefunctions in the various
regions of the ring in absence of magnetic flux are given as follows
\begin{eqnarray}
\psi_0&=&  e^{ikx_{0}}+r e^{-ikx_{0}},\nonumber\\
\psi_1&=&a e^{ikx_{1}}+b e^{-ikx_{1}},\nonumber\\
\psi_2&=&c e^{ikx_{2}}+d e^{-ikx_{2}},\nonumber\\
\psi_3&=&e e^{ikx_{3}}+f e^{-ikx_{3}},\nonumber\\
\psi_4&=&g e^{ikx_{4}}+h e^{-ikx_{4}}.
\end{eqnarray} 

Here $x_{i},i=0,..4$ are coordinates along the connecting lead to the
reservoir, and the segments J1J2 , J2bJ3, J2aJ3, and J3J1
respectively.  The Fermi wavevector is defined as
$k=\sqrt{2mE/\hbar^2}$. To solve for the unknown coefficients in
eqn.(1) we use Griffith\cite{xia} boundary condition at the
junctions $J1, J2$ and $J3$. These boundary conditions are due to the
single-valuedness of wavefunction and current conservation (Kirchoff's
law). In the presence of magnetic flux in the system we can choose a
gauge for the vector potential in which the field does not appear
explicitly in the Hamiltonian. The boundary conditions do not change,
however the electron propagating from one junction to another picks up
an additional phase, which is positive for clockwise motion and
negative for anti-clockwise motion, but of same magnitude. For further
details see Refs.[\onlinecite{prb1994,gia}].  Naturally different
segments pick up different phases. Using the above mentioned boundary
conditions we get

\begin{eqnarray}
1 + r = a + b e^{-i\alpha_{1}}= g e^{ikl_{4}+ i\alpha_{4}} + h e^{-i
k l_{4}},\nonumber\\
1 - r - a +b e^{-i\alpha_{1}}+  g e^{ikl_{4}+ i\alpha_{4}}-h e^ {-i
k l_{4}} = 0 ,\nonumber\\  
a e^{ikl_{1}+ i\alpha_{1}}+ b e^{i k l_{1}}= c + d
e^{i\alpha_{2}}= e + f e^{i\alpha{3}},\nonumber\\
a e^{ikl_{1}+ i\alpha_{1}} - b e^ {-i k l_{1}} - c + d
e^{-i\alpha_{2}} - e + f e^{-i\alpha_{3}} = 0,\nonumber\\
c e^{ikl_{2}+i\alpha_{2}} + d e^ {-i k l_{2}} = e e^{ikl_{3}+
i\alpha_{3}}+ f e^{- ikl_{3}} = g + h e^{- i\alpha_{4}},\nonumber\\
c e^{ikl_{2}+ i\alpha_{2}} - d e^{-i k l_{2}} + e e^{ikl_{3}+
i\alpha_{3}} - f e^{-ikl_{3}} - g + h e^{-i\alpha_{4}}= 0. 
\end{eqnarray}

  Here $\alpha_{1},\alpha_{2},\alpha_{3}$ and $\alpha_{4}$ are phases
  picked up by the wavefunctions in the segments J1J2, J2bJ3, J2aJ3
  and J3J1 respectively and we have
  $\alpha_{1}+\alpha_{2}+\alpha_{4}=2\pi\phi/\phi_{0},$ and
  $\alpha_{1}+\alpha_{3}+\alpha_{4}=2\pi\phi/\phi_{0}$, such that
  $\alpha_{2}=\alpha_{3}$ as required by definition. Using eqn.(2) we
  have solved for all the unknown coefficients in eqn.(1).
  
  In the lead connecting the reservoir to our circuit there is no
  current flow as $|r|^2=1$. Throughout the discussion the lengths are
  scaled with respect to the total length of the bubble
  $l=l_{2}+l_{3}$. The wavevector $k$ is identified in a dimensionless
  form $k \equiv kl$.  The probability current density is defined as
  $J=\frac{e\hbar}{2mi}(\psi^*\nabla\psi-\psi\nabla\psi^*)$. For the circuit
  segment $J1J2$ of the figure (1), when we derive the probability current
  density  we get-
  $J=\frac{e\hbar k}{m}(|a|^2-|b|^2)$. Now the current densities $(I)$ in
  their dimensionless form are given by dividing $J$ by
  $\frac{e\hbar k}{m}$.  This approach is widely
  used in  literature to define the current densities, see
  Refs.[\onlinecite{buti,physicapareek}]. The current
  densities in the small
  interval $dk$ around the Fermi energy $k$ in the various segments of
  the circuit are given by -
\begin{eqnarray}
I_{1}=|a|^2-|b|^2,\nonumber\\
I_{2}=|c|^2-|d|^2,\nonumber\\
I_{3}=|e|^2-|f|^2,\nonumber\\
I_{4}=|g|^2-|h|^2.
\end{eqnarray}

Just to mention again that $I_{1},I_{2},I_{3}$ and $I_{4}$ are the
persistent current densities in the segments $J1J2, J2bJ3, J2aJ3$ and
$J3J1$ respectively. The persistent current densities in various parts
of the circuit show cyclic variation with flux and $\phi_{0}$
periodicity, and oscillate between positive and negative values as a
function of energy or the wavevector $k$ as expected.  Since the
analytical expressions for these currents are too lengthy we confine
ourselves to a graphical interpretation of the results. It should be
noted that in all these currents flux enters only through the
combinations $\alpha_{1}+\alpha_{2}+\alpha_{4}$ and
$\alpha_{1}+\alpha_{3}+\alpha_{4}$ the magnitude of these combinations
is given by $2\pi\phi/\phi_{0}$ as expected. For us the current
densities in the bubble(J2bJ3aJ2) are of special importance as in this
region there is a possibility of current magnification which will be
analysed below.  The currents induced in segment J3J1 and J1J2 are
equal, i.e $I_{1}=I_{4}$. These currents may have positive(clockwise)
or negative(anticlockwise) values depending on the flux $\phi$ and
value of Fermi wavevector $k$. For a fixed $k$ this current oscillates
between positive and negative values as a function of $\phi$ with a
period $\phi_{0}$ and are asymmetric in $\phi$. Similarly for fixed
value of $\phi$ currents oscillate as one varies $k$. The magnitude of
current shows a maximum or minimum near the corresponding eigen-states
of the system. We have calculated these
eigen states for two different cases. For open system as depicted in
figure (1) one can calculate the energies(or wavevector) of these
states by looking at the poles of the S-Matrix. These states
correspond directly to resonances. In our case S-Matrix
is simply a complex reflection amplitude $r$.  We have also analysed
the eigen states of a closed system(without coupling lead to
reservoir) by wavefunction matching in various segments using
waveguide theory. The eigenvalues are obtained by solving the
following equation, resulting from waveguide theory,

\begin{eqnarray}
 \cos(\alpha) & = & \frac{1}{\cos(kl_{-})} (\cos k(l_{1}+l_{+}) - \nonumber\\
              &   & \frac{1}{4} \frac{\sin (kl_{1}) \sin (kl_{2})
                    \sin (kl_{3})}{\sin (kl_{+})}),
\end{eqnarray}

where, $\alpha=2\pi\phi/\phi_{0}$, $l_{+}=(l_{2}+l_{3})/2$ and
$l_{-}=(l_{2}-l_{3})/2$.

\begin{figure}
\protect\centerline{\epsfxsize=3.0in\epsfbox{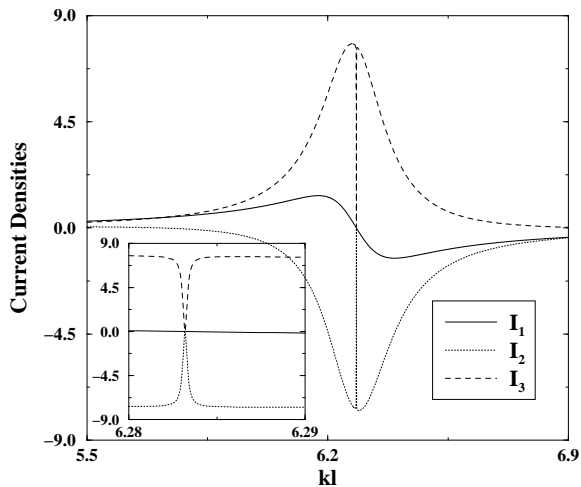}}
\caption{Persistent current densities are shown as function of
  $kl$. The lengths are $l_{1}/l=l_{4}/l=0.25,l_{2}/l=
  0.45,l_{3}/l=0.55$ and flux $\phi=0.1$. In the inset we have shown
  the current densities around the value wherein $I_{1}$ goes to zero.}
\end{figure}
We analyse the case of a bubble with unequal lengths, of its two arms,
i.e., the length of $J2bJ3$ $\neq$ $J2aJ3$. This asymmetry implies
that current densities in the two arms of the bubble $I_{2}$ $\neq$
$I_{3}$. In figure (2), we plot the persistent current densities in
various parts of the circuit.  It should be noted that absolute value
of the persistent current densities $I_{2}$ and $I_{3}$ are
individually much larger than the input current density $I_{1}$ into
the bubble and thus the current magnification effect is
evident(without violating the basic Kirchoff's law). The input current
arises due to the presence of flux $\phi$ as it breaks the time
reversal symmetry. The physical parameters used for this figure are
mentioned in the figure caption. In the interval $5.5<kl<6.9$ the
current $I_{1}$ changes from positive to negative and exhibits
extremum around the real part of the poles of the S-Matrix($6.278$ and
$6.328$). For the closed system the eigen values are at $5.93$ and
$6.68$. The difference between eigenvalues for closed and open
systems(quasi bound states) arise from the additional scattering from
the junction $J1$ coupled to the reservoir. Moreover, eigenvalues for
open systems are complex, as electron has a finite lifetime in the
ring system before entering into the reservoir. When $I_{1}$ is
positive, negative current density of magnitude $I_{2}$ flows in the
arm $J2bJ3$ of the bubble.  Thus, when $I_{1}$ is positive circulating
current flows in the anti-clockwise direction in the bubble. In the
range where $I_{1}$ is negative, i.e, input current into the bubble is
in anticlockwise direction, then positive current flows in arm
$J2aJ3$. According, to our convention as mentioned earlier,
circulating current flows in the anti-clockwise direction.  The
magnitude of this circulating current density $I_{c}$, is taken to be
the value of current in one of the arms of the bubble moving against
the input current into the bubble as explained in detail in the
introduction.

\begin{figure}
\protect\centerline{\epsfxsize=3.0in\epsfbox{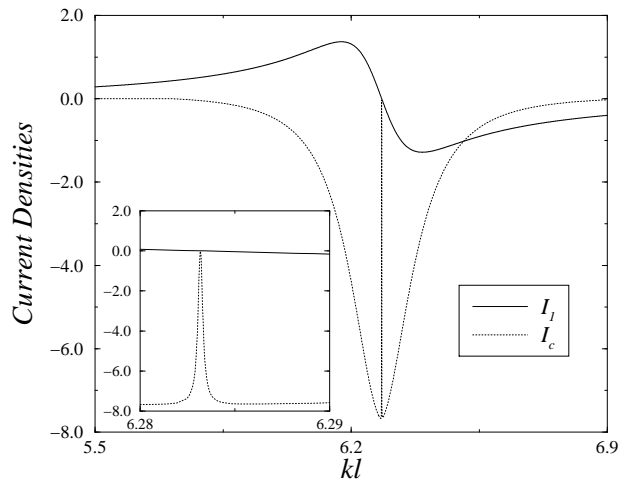}}
\caption{Persistent current density $I_{1}$ and circulating  current
  density $I_{c}$ is plotted as function of $kl$. The parameters are
  same as used in fig.~2. The inset shows the behavior of $I_{c}$ and 
  $I_{1}$ around their zero values.}
\end{figure}

In figure (3) we have plotted the persistent current density
$I_{1}=I_{4}$ and the circulating current density $I_{c}$ in the
bubble for the same parameters used in figure~2. It should be noted
that if we interchange the values of $l_{2}$ and $l_{3}$ keeping other
parameters unchanged circulating current will flow in a clockwise
direction. This is obvious from the geometry of the problem.

We generally observe current magnification at those Fermi energy
wavevector intervals around the eigen energies of the
system\cite{prb1994,physicapareek}. However, there are some
exceptions. In figure (4), we plot one of those exceptions. The new
physical parameters are mentioned in the figure caption. In figure(4)
we show that current magnification does not occur at places which are
eigen values of the aforesaid system. Here the real part of the eigen
wavevector $kl$ corresponds to $10.184$ (for closed system it is at
$10.171$). One can readily notice that the magnitude of persistent
current (i.e, input current $I_{1}$) shows extrema around this value.
Around this region both the currents in the bubble $I_{2}$ and $I_{3}$
are individually smaller than $I_{1}$ and they flow in the same
direction as the input current.  Hence we do not observe current
magnification effect around this quasi bound state of the open system.
We also observe that current magnification does occur at some places
which are not near the eigen values of the system.
\begin{figure}
\protect\centerline{\epsfxsize=3.0in\epsfbox{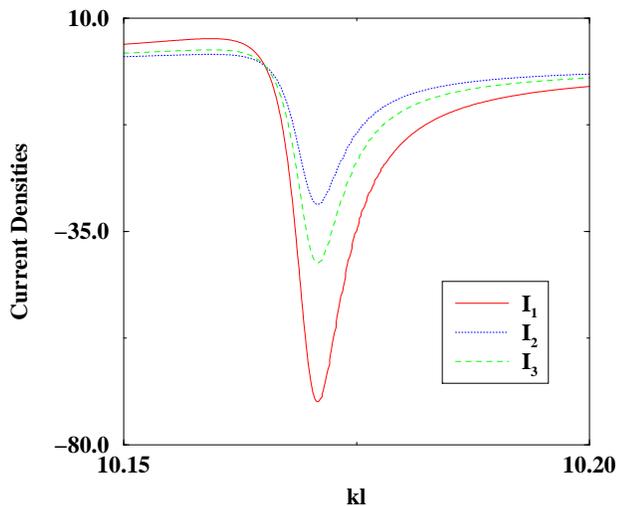}}
\caption{Persistent current densities are platted as function of
  $kl$. The lengths are $l_{1}/l=l_{4}/l=0.25,l_{2}/l=
  0.15,l_{3}/l=0.85$. Flux $\phi=0.1$. }
\end{figure}

All these figures establish the fact that the current magnification
effect (and associated circulating currents) which are quantum
mechanical in origin are extremely sensitive to the system parameters.
The exact conditions for current magnification cannot be readily
predicted \emph{a priori}. The orbital magnetic moment of the system
is given by the line integral of the total current taken across the
entire system. The total current is given by integrating the current
densities upto the Fermi energy (at temperature $T=0$). If the system
exhibits current magnification effect one should be able to detect it
experimentally by observing the enhanced response of the magnetic
moment by appropriate tuning of Fermi energies. We expect systems
comprising several metallic loops interwoven together to exhibit a new
feature in the magnetic response due to current magnification. It
should be noted if the whole system is embedded in a magnetic field
then we have both persistent currents as well as circulating currents
that can be separated by their symmetry properties under flux
reversal\cite{physicapareek}. Just for the sake of simplicity and to
show the existence of current magnification in equilibrium we have
taken a system in which bubble does not enclose a magnetic flux, which
may not be an ideal system. However, it clarifies our contention.

In conclusion we have shown that current magnification effect can
occur in equilibrium mesoscopic systems in presence of magnetic flux.
Earlier, it was shown to occur in a non-equilibrium
state\cite{prb1994}. This quantum effect is extremely sensitive to
system parameters. Our system also exhibits breakdown of parity
effects (using eqn. (4)) \cite{cheung2}.  This, along with analysis of
current magnification in presence of magnetic flux, encompassing the
entire sample will be reported elsewhere.

\end{document}